\documentclass{aa}
\usepackage{graphicx}
\usepackage{epsfig}
\usepackage{txfonts}
\usepackage{natbib}
\begin{document}

\title{Ideal kink instability of a magnetic loop equilibrium}

\author{T. T\"or\"ok\inst{1,2}
   \and B. Kliem    \inst{1}
   \and V. S. Titov \inst{3}}

\institute{Astrophysikalisches Institut Potsdam, 14482~Potsdam, Germany
      \and School of Mathematics and Statistics, University of
           St\,Andrews, St\,Andrews, Fife KY16 9SS, UK
      \and Theoretische Physik IV, Ruhr-Universit\"at Bochum, 44780 Bochum,
           Germany}


\offprints{T. T\"or\"ok, \email{ttoeroek@aip.de}}

\date{Received 25 September 2003 / Accepted 3 November 2003 } 

\abstract{The force-free coronal loop model by \cite{Tit:Dem-99} is found
to be unstable with respect to the ideal kink mode,
which suggests this instability as a mechanism for the initiation of flares.
The long-wavelength ($m\!=\!1$) mode grows for average
twists $\Phi\ga3.5\pi$ (at a loop aspect ratio of $\approx$\,5).
The threshold of instability increases with increasing
major loop radius, primarily because the aspect ratio then also increases.
Numerically obtained equilibria at subcritical twist are very close to the
approximate analytical equilibrium; they do not show indications of
sigmoidal shape.
The growth of kink perturbations is eventually slowed down by the surrounding
potential field, which varies only slowly with radius in the model. With
this field
a global eruption is not obtained in the ideal MHD limit. Kink
perturbations with a rising loop apex lead to the 
formation of a vertical current
sheet below the apex, which does not occur 
in the cylindrical approximation. 

}

\maketitle
\section{Introduction}
\label{intro}
Magnetic loops are the elementary building block of the solar corona and of
low-plasma-beta environments of astrophysical objects in general.
Understanding their instabilities is one of the fundamental problems in
corona physics. Due to the small plasma beta
($\beta=P_\mathrm{kin}/P_\mathrm{mag}\sim10^{-3}\dots10^{-2}$ in the inner
solar corona), the magnetic configuration of stable loops must be nearly
force free. Any instability must be caused by currents flowing mainly along
the loops. A sigmoidal (S- or inverse-S) shape of the projection of loops
onto the solar surface is regarded as a signature of such currents, and
indeed, a strong sigmoidal shape of soft X-ray loops correlates with
eruptive activity \citep{Canf:al-99}. 

The stability of current-carrying force-free (or nearly force-free) fields
was extensively studied for toroidal geometry in fusion research \cite[see,
e.g.,][]{Biskamp-93} and for cylindrical geometry with fixed ends in the
astrophysical context. The latter is considered as an approximation to
coronal loops with large aspect ratio which includes the effect of
photospheric line tying \cite[e.g.,][]{Hoo:Pri-81,
Vell:al-90,Miki:al-90,Einaudi-90,Hood-92,Bat:Hey-96,Baty-01,Gerr:al-02}.
It was found that the stability is mainly controlled by the total twist,
\begin{equation}
\Phi = \frac{lB_\phi(r)}{rB_z(r)}\,,
\label{eq_Phi}
\end{equation}
where $l$ is the length of the current-carrying flux system, $r$ is
the (minor) radius, and $B_z$ and $B_\phi$ are the axial and azimuthal
field components, respectively. In addition, the radial profile of the
twist, the effect of line tying, and the ratio of radius and twist scale
length ($\lambda=2{\pi}l/\Phi$) are important. For example, the
uniformly twisted force-free toroidal or periodic cylindrical configuration
is kink-unstable for $\Phi>\Phi_\mathrm{c}=2\pi$, but line tying raises the
instability threshold to $\approx2.5\pi$ \citep{Hoo:Pri-81,Ein:VHov-83}.
Localizing the current within a certain radius, $a$, and embedding it
in a potential field also raises the threshold
(as any substantial radial variation of the twist). Configurations of this
type were found to be stable for $\Phi_\mathrm{max}\la5\pi$, where
$\Phi_\mathrm{max}$ is the peak value of the twist in the configuration
\citep{Miki:al-90,Bat:Hey-96}.
The instability threshold of embedded loops rises strongly if their radius
becomes very small, i.e. if $\delta=2{\pi}a/\lambda\la1$ \citep{Baty-01}.
For cylindrical line-tied force-free configurations it was also shown that
the longest-wavelength kink mode (azimuthal wave number $m\!=\!1$) becomes
unstable before any other mode in the ideal MHD limit \citep{VdLin:Hoo-99}. 

The stability of arched magnetic loops has not yet been investigated.
However, numerical studies of the injection of twist by slow vortex motions
at the footpoints of an initially current-free loop-shaped flux bundle
indicate that raising the twist of a loop beyond a critical value leads to
destabilization \citep{Amar:al-96,Toe:Kli-03}. The critical twist lies in
the range $2.5\pi<\Phi_\mathrm{c}<2.75\pi$ for the specific initial
configurations studied. The driving by photospheric vortices generally
influences a large surrounding volume in addition to the twisted loop.
It also causes an increase of the loop length and width with increasing
twist. Clearly, a stability consideration of loops free from such
additional influences is needed. 

In this paper we study the stability of the loop model by
\cite{Tit:Dem-99}, cited as T\&D in the following
(see their Fig.\,2 for a schematic). This approximate,
cylindrically symmetric, force-free equilibrium consists of a toroidal ring
current of major radius $R$ and minor radius $a$, whose outward-directed
Lorentz self-force is balanced with the help of a field by two fictitious
magnetic charges of opposite sign which are placed at the symmetry axis of
the torus at distances ${\pm}L$ to the torus plane. That axis lies below
the photospheric plane $\{z\!=\!0\}$ at a depth $d$. The resulting field
outside the torus is current-free and contains a concentric magnetic X line
between the torus and its centre. A toroidal field component created by a
fictitious line current running along the symmetry axis is added.
Its strength controls the twist of the
field in the torus, and it turns the X line into a hyperbolic flux tube
(HFT; for a definition, see \citeauthor{Tito:al-02},
\citeyear{Tito:al-02}). The existence of the HFT is generic to
such force-free configurations with a net current. The accuracy of the
obtained equilibrium improves with decreasing parameters $a/R$ and $a/L$.
Its section in the volume $\{z\!>\!0\}$ is a model of a coronal magnetic
loop.

T\&D have investigated the stability of the torus with respect to global
expansion (growing perturbations ${\delta}R>0$) and found instability for
sufficiently large radii, $R\ga\sqrt{2}\,L$.
\cite{Rous:al-03} confirmed this recently by numerical simulations, which
also suggested that the instability threshold lies near $5L$ and that the
slow decrease of the surrounding toroidal field with distance from the
generating line current leads to saturation of the instability and prevents
the loop from erupting.
In the following, we investigate the stability of the
configuration with respect to ideal kink modes in dependence of the twist
in its coronal part. 
\section{Numerical model}
\label{numerics}
Using the force-free equilibrium by T\&D as the initial condition, we
integrate the compressible ideal MHD equations to study whether
instabilities occur. Based on the small value of the plasma beta in the
inner corona, we use the simplifying assumption $\beta=0$ in most of our
parametric study: 
\begin{eqnarray}
\partial_t\rho&=&
                 -\nabla\cdot(\rho\,\vec{u})\,,           \label{eq_rho}\\
\rho\,\partial_{t}\vec{u}&=&
      -\rho\,(\,\vec{u}\cdot\nabla\,)\,\vec{u}
      +\vec{j}\mbox{\boldmath$\times$}\vec{B} 
      +\nabla\,\cdot\tens{T}\,,                           \label{eq_mot}\\
\partial_{t}\vec{B}&=& 
    \nabla\mbox{\boldmath$\times$}(\,\vec{u}\mbox{\boldmath$\times$}
    \vec{B}\,)\,,                                         \label{eq_ind}\\
\vec{j}&=&\mu_0^{-1}\nabla\mbox{\boldmath$\times$}\vec{B}\,. \label{eq_cur}
\end{eqnarray}
Here $\tens{T}$ denotes the viscous stress tensor 
($\tens{T}_{ij}=\rho\,\nu\,[\partial u_{i}/\partial x_j+
                            \partial u_{j}/\partial x_i-
                       (2/3)\delta_{ij}\,\nabla\cdot\vec{u}]$)
and $\nu$ is the
kinematic viscosity, included to facilitate relaxation toward equilibrium
in the stable cases. 

The condition $\beta=0$ ensures that the numerically obtained stable
equilibria are force free, and in the unstable regime it yields higher
growth rates than the runs with $\beta>0$.
Hence, the threshold for onset of instability is correctly obtained. For a
few parameter sets, growth rates were also determined with the pressure
gradient term, $-\nabla{p}$, included in Eq.\,(\ref{eq_mot}); a standard
form of the energy equation and an adiabatic equation of state were then
added to Eqs.\,(\ref{eq_rho}-\ref{eq_cur}), as e.g., in \cite{Klie:al-00}. 

The equations are normalized in the usual manner \cite[see,
e.g.,][]{Toe:Kli-03} by quantites derived from a characteristic length,
taken here to be $L$, and the magnetic field, $B_0(0,0,R\!-\!d)$, and
Alfv\'en velocity,
$v_\mathrm{a0}(0,0,R\!-\!d)$, at the loop axis at $t\!=\!0$.
The loop is chosen to lie in the plane $\{x\!=\!0\}$. The initial density
distribution is specified such that the Alfv\'en velocity is uniform,
$\rho_0=B_0^2$, and in most cases the computation is started with the
system at rest, $\vec{u}_0=0$. 

A modified
Lax-Wendroff scheme is used on a nonuniform Cartesian grid in a box
$[-L_x,L_x]\times[0,L_y]\times[0,L_z]$
with minimum grid spacing at the origin and very slowly increasing grid
spacing toward the upper, side, and back boundaries.
$L_x\!=\!L_y\!=\!L_z/2\!=\!5$ and
$\Delta x_{\rm min}\!=\!\Delta y_{\rm min}\!=\!\Delta z_{\rm min}\!=\!0.02$
are used. Line symmetry with respect to the $z$ axis is prescribed at the
front boundary, $\{y\!=\!0\}$, in order to use the available computing
resources efficiently for maximum resolution. This implies
$u_{x,y}(0,0,z)=0$. The variables are held fixed at their initial values at
all other boundaries except for the bottom, where the tangential
magnetic field components are extrapolated onto the ghost points.
This permits a weak evolution of the fields in the plane
$\{z\!=\!0\}$ ($\vec{u}(t)\!=\!0$ and $\rho(t)\!=\!\rho_0$ in
$\{z\!<\!0\}$), enhancing stability in comparison to a bottom boundary with
fixed variables (i.e., permitting the use of small diffusion parameters). A
test run with fixed magnetic field at the bottom boundary (at $R=2.2$,
$\Phi_\mathrm{loop}=4.9\pi$) gave qualitatively identical results with a
reduction of the growth rate by only 4 percent. A small amount of
artificial spatial smoothing, $1-\sigma_\rho=1-\sigma_u=0.005$, is applied
to the variables $\rho$ and $\vec{u}$, respectively, to stabilize the
scheme in addition to the stabilization by the viscosity ($\nu=0.05$); see
\cite{Toe:Kli-03} for a detailed description of the numerical tools.  
\section{Kink instability}
\label{instab}
\begin{figure}    
\resizebox{\hsize}{!}                
          {\includegraphics{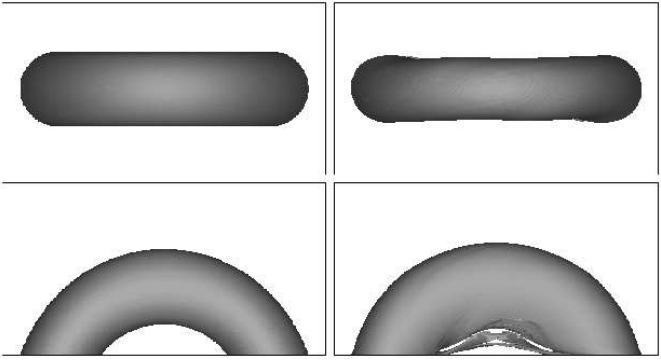}}
\caption[]
{Top and side view of the isosurface $|\vec{j}|=0.4\,j_{\rm max}$ for
 $\Phi_\mathrm{loop}\!=\!2.1\pi$, $R\!=\!2.2$, $\beta=0$ at $t\!=\!0$
 (\emph{left}) and
 $t\!=\!78$ (\emph{right}). Times are in units of $\tau_\mathrm{a}$. 
 The volume $|x|\le1.5$, $|y|\le3$, $0\le z\le3$ is shown.} 
\label{fig_iso_j_Nt=5}
\end{figure}

We start with a parameter set that is very close to those
used by T\&D and yields a
configuration with left-handed field line twist
which is stable against both global expansion and kink modes.
We choose $d=L=1$ (50~Mm), $R=2.2$, and 
the number of field line turns about the axis of the whole torus at its
surface to be $N_\mathrm{t}=5$, the coronal part being
$N_\mathrm{loop}=1.75$. The values of the currents and charges used by T\&D
and their equilibrium conditions
then yield $a=0.65$. (The larger $R$, the smaller the resulting $a/R$, so
we have chosen a value of $R$ not much smaller than the estimated stability
limit of the
global expansion mode). The resulting coronal loop is shown in
Fig.\,\ref{fig_iso_j_Nt=5}. 

\begin{figure}    
\begin{center}
\resizebox{.86\hsize}{!}                
          {\includegraphics{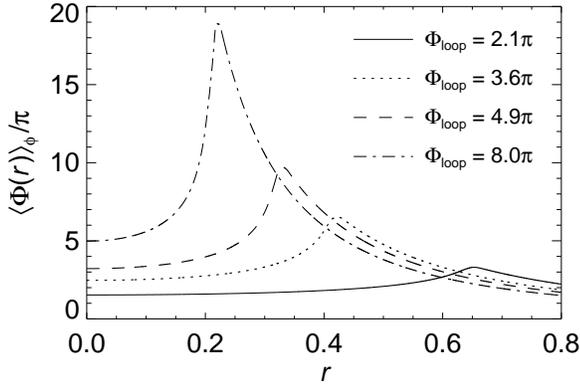}}
\end{center}
\vspace{-3mm}
\caption[]
{Radial profile of the twist at the loop apex, averaged over azimuth angle
 $\phi$; 
 $\langle\Phi(r)\rangle_\phi=
         (l/2\pi r)\int(B_\phi(r,\phi)/B_y(r,\phi))\mathrm{d}\phi$, 
 $l=2R\arccos(d/R)$, 
 $r=(x^2+(z-R+d)^2)^{1/2}$, $\phi=\arctan((z-R+d)/x)$; 
 $\Phi_\mathrm{loop}=
  (2/a^2)\int_0^a\langle\Phi(r)\rangle_\phi r\mathrm{d}r$.}
\label{fig_phig}
\end{figure}

The T\&D equilibrium has the property that the radial variation of the twist
becomes substantial for $N_{\mathrm{t}}{\ga}R/a$ (Fig.\,\ref{fig_phig}).
The average twist of this configuration is found to be
$\Phi_\mathrm{t}=6.0\pi$, the fraction in its coronal part being
$\Phi_\mathrm{loop}=2.1\pi$ (Table\,\ref{Tab1}).
Figure\,\ref{fig_iso_j_Nt=5} shows that the loop is stable for these
parameters and relaxes to a numerically force-free configuration very close
to the approximate analytical equilibrium.  The fluid velocity at the loop
apex oscillates about zero with monotonically decreasing amplitude
($<9\times10^{-5}$ by $t\!=\!78$). The HFT collapses into a small vertical
current sheet as the loop settles to the numerical equilibrium, causing its
inner edge to bulge out slightly (both effects are very weak at higher
$\Phi$, as long as $\Phi_\mathrm{loop}<\Phi_\mathrm{c}$.) 

To check the stability of the loop, we have applied velocity perturbations
uniform in direction with peak magnitude $u_1\!=\!0.05$ in a spherical
volume with radius $a$ and Gaussian profile at the loop apex, ramped up
over 10 Alfv\'en times ($\tau_\mathrm{a}\!=\!L/v_\mathrm{a0}$) and then
switched off. Both upward and downward directed perturbations were damped
away within $\sim20\tau_\mathrm{a}$. 
%

\begin{figure}    
\resizebox{\hsize}{!}                
          {\includegraphics{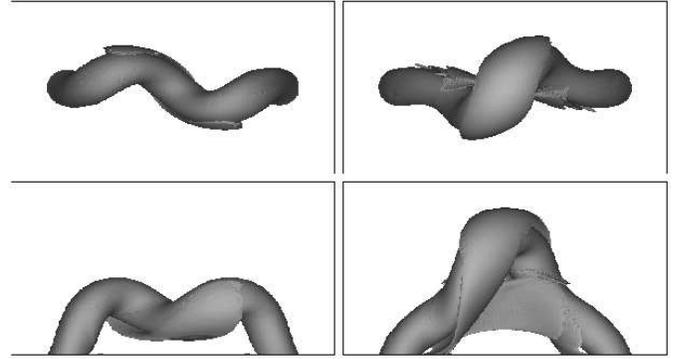}}
\caption[]
{Top and side view of current density isosurfaces for
 $\Phi_\mathrm{loop}\!=\!4.9\pi$, $R\!=\!2.2$, $\beta\!=\!0$.
 \emph{Left}:  $|\vec{j}|=0.15\,j_{\rm max}$ at $t\!=\!35$; unperturbed case.
 \emph{Right}: $|\vec{j}|=0.25\,j_{\rm max}$ at $t\!=\!28$; run with an
                upward initial velocity perturbation at the loop apex;
               see the online edition for an animation of these data.
 $|x|\le1.5$, $|y|\le3$, $0\le z\le3$.} 
\label{fig_iso_j_Nt=15}
\end{figure}

\begin{figure}    
\begin{center}
\resizebox{.9\hsize}{!}                
          {\includegraphics{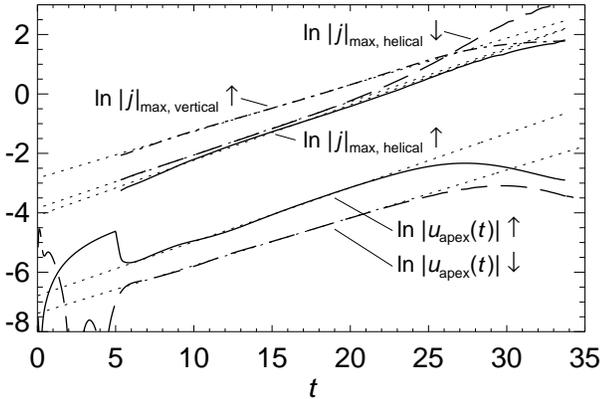}}
\end{center}
\vspace{-3mm}
\caption[]
{Apex velocity and peak current density, $|j(0,0,z,t)|_\mathrm{max}$, in
the current sheets for upward ($\uparrow$) and downward ($\downarrow$) apex
displacements. $R\!=\!2.2$, $\Phi_\mathrm{loop}\!=\!4.9\pi$, $\beta\!=\!0$.
Exponential fits are shown dotted.}
\label{fig_growth}
\end{figure}

Next we consider the case $N_\mathrm{t}=15$ ($\Phi_\mathrm{loop}=4.9\pi$).
This configuration shows the spontaneous
development of the long-wavelength ($m\!=\!1$)
kink mode very clearly (Figs.\,\ref{fig_iso_j_Nt=15} and \ref{fig_growth}).
The mode is initiated
by the weak, downward-directed forces that result from the initial
discretization errors of the current density
on the grid. The figures also show a run with a small
upward initial velocity perturbation applied at the apex ($u_1=0.01$,
ramped up over $5\tau_\mathrm{a}$); in this case the $m\!=\!1$ kink mode
grows in a similar manner in the linear phase, with an azimuthal phase
shift of $\pi$ (opposite sigmoidal shape). Similar to the cylindrical case,
the growth of the mode leads to the formation of a helically shaped current
sheet at the interface of the perturbed loop with the surrounding
magnetofluid. The upward directed kink instability of the loop forms a
second, nearly vertical
current sheet, at the HFT underneath, which does not occur in
the cylindrical kink.
The peak current density in these sheets
rises exponentially in the linear phase of the instability and exceeds that
in the kinked loop in the nonlinear stage.
The implications of the helical kinking and current sheet formation for the
interpretation of sigmoidal structures in the solar corona have been
discussed in \cite{Klie:al-03}. 
%
\begin{table}[t]
\caption{Normalized parameters of the runs with $d\!=\!L\!=\!1$ and
$\beta\!=\!0$, and growth rate of the kink instability
for upward and downward apex motion.
$\delta=(a/l)\Phi_\mathrm{loop}$. Growth rates marked with a dag refer to 
runs with an initial velocity perturbation. The parameters of the first run
correspond to the values given in Fig.\,4 of T\&D (except $R$).} 
\begin{tabular}{ccccccc}
$R$ & $N_\mathrm{t}$ 
           & $a$ & $\Phi_\mathrm{loop}/\pi$ 
	                & $\delta$ & 
                               ${\gamma_\mathrm{up}}a/v_\mathrm{a0}$ & 
			          ${\gamma_\mathrm{down}}a/v_\mathrm{a0}$ \\
\hline
2.2 & ~5   & 0.65 & 2.1 & 0.88 & 0             & 0            \\
2.2 & ~9   & 0.44 & 3.3 & 0.96 & 0             & 0            \\
2.2 & 10   & 0.42 & 3.6 & 0.98 & 0.008$^\dag$  & 0.001$^\dag$ \\
2.2 & 12   & 0.37 & 4.2 & 1.00 & 0.031$^\dag$  & 0.023        \\
2.2 & 15   & 0.32 & 4.9 & 1.03 & 0.059$^\dag$  & 0.052        \\
2.2 & 20   & 0.27 & 6.0 & 1.06 & 0.088$^\dag$  & 0.076        \\
2.2 & 25   & 0.24 & 7.0 & 1.08 & 0.102         & 0.091$^\dag$ \\
2.2 & 30   & 0.21 & 8.0 & 1.10 & 0.111         & 0.100$^\dag$ \\
3.4 & ~9.1 & 0.44 & 5.0 & 0.80 & 0.005         &              \\
3.4 & 11.4 & 0.38 & 6.0 & 0.83 & 0.021         &              \\
3.4 & 13.9 & 0.34 & 7.0 & 0.87 & 0.034         &              \\
3.4 & 16.4 & 0.31 & 8.0 & 0.89 & 0.044         & 
\end{tabular}
\label{Tab1}
\end{table}
%
Figure\,\ref{fig_growth} shows that the apex displacement grows
exponentially: an instability occurs.
The total kinetic energy in the box grows exponentially as well.
Upward perturbations grow at a
slightly higher rate because the restoring forces due to the toroidal field
by the line current, the inertia of the fluid, and the effect of approach
to a closed boundary are all weaker than for downward displacements. We
expect that the kink mode with a sideward directed perturbation at the loop
apex, which can in a first approximation be regarded as an azimuthally
phase-shifted mode, is quantitatively similar to the modes with vertical
perturbation at the apex studied here. 

\begin{figure}    
\begin{center}
\resizebox{.9\hsize}{!}                
          {\includegraphics{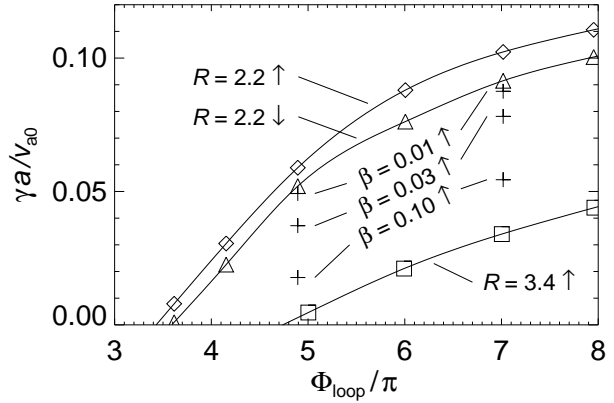}}
\end{center}
\vspace{-3mm}
\caption[]
{Kink instability growth rates of the T\&D equilibrium of radius $R$ derived
 from the upward ($\uparrow$) and downward ($\downarrow$) apex velocities.
 Open symbols refer to runs with $\beta\!=\!0$. $R\!=\!2.2$ for all
 $\beta\!>\!0$ runs.}
\label{fig_gamma}
\end{figure}

A fit of the growth rates for $R=2.2$ suggests a critical average
twist for onset of kink instability in the T\&D equilibrium of
$\Phi_\mathrm{c}\approx3.5\pi$, occurring at an aspect ratio of
$\approx$\,5 (Fig.\,\ref{fig_gamma}, Table\,\ref{Tab1}).
This threshold is similar to that found for cylindrical equilibria with
similar twist profiles \cite[e.g.,][]{Miki:al-90,Bat:Hey-96}. The growth
rates, when normalized in the same manner
(${\gamma}a/v_\mathrm{a0}[\Phi_\mathrm{loop}-\Phi_\mathrm{c}]$), are also
similar (the factor $\approx$\,1/2 in our growth rates nearly disappears
if viscosities $\nu\!\la\!0.005$ are chosen; however, then
the $m\!=\!2$ mode starts to grow, which will be investigated separately).

A similar picture is obtained for larger loop radius $R$ (and aspect ratio
$R/a$). Since the parameter $\delta$ decreases with increasing $R$, the
threshold of instability increases (Fig.\,\ref{fig_gamma}). The threshold
is probably smaller than $\approx$\,3.5$\pi$ for $R<2.2$, but the
corresponding small
aspect ratios are not characteristic of solar coronal loops. 
All growth rates are invariant against a reversal of the sign of the twist. 

Runs with nonvanishing pressure (set uniform initially such that at the
loop axis $\beta=0.01;\,0.03;\,0.1$) show identical
qualitative behaviour with reduced growth rates (Fig.\,\ref{fig_gamma}). 

The instability enters a nonlinear saturation phase in all runs. The apex
velocity then drops, but the current density in the
formed sheets continues to rise until numerical instability is unavoidable.
Apparently, a global eruption is not reached in ideal-MHD simulations, due
to the counteraction by the strong overlying field in the model, as 
also found by \citeauthor{Rous:al-03}
However, the exponential rise and helical shape of the loop and the
saturation correspond excellently to eruptive filaments that do not succeed
to escape from the Sun \citep{Ji:al-03}. 
%

Finally, we have checked whether the numerical equilibria develop a
sigmoidal shape as the twist approaches $\Phi_\mathrm{c}$ from below. No
indication of sigmoidality could be found up to $\Phi_\mathrm{loop}=3.3\pi$
($N_t=9$). Even after applying an upward initial perturbation as in the run
with $N_t=5$, the loop relaxed to its original circular shape, straight in
projection and similar to that shown in Fig.\,\ref{fig_iso_j_Nt=5}. 
\section{Conclusions}
\label{concl} 
The magnetic loop equilibrium by \cite{Tit:Dem-99} is
kink-unstable for twists $\Phi>\Phi_\mathrm{c}$, with
$\Phi_\mathrm{c}\approx3.5\pi$ at a loop aspect ratio $R/a\approx5$. The
instability threshold rises with rising aspect ratio. No indication of
sigmoidal shape at slightly subcritical twists was found. The unstable kink
mode with upward displacement of the loop apex forms a vertical current
sheet at the HFT
underneath the loop, which has no counterpart in kink-unstable cylindrical
loop models
but corresponds to the central element of the ``standard flare model''
\cite[e.g.,][]{Shibata-99}. The surrounding potential field eventually
terminates the exponential growth of
kink modes in the considered configuration, apparently
inhibiting a global
eruption within the framework of ideal MHD. However, the configuration may
nevertheless suffer a global eruption if (a) a surrounding field which
decreases more rapidly with height is employed, (b) magnetic reconnection
is permitted to occur in the formed current sheets,
or (c) magnetic reconnection with neighbouring
loops is triggered in a multiple-loop configuration. 
\begin{acknowledgements}
This investigation was supported by BMBF/DLR grants 50\,OC\,9901\,2,
and 01\,OC\,9706\,4, by the Volkswagen Foun\-da\-tion, and by EU grant
HPRN-CT-2000-00153. The John von Neumann-Institut f\"ur Computing, J\"ulich
granted Cray computer time.
\end{acknowledgements}
%
%
%
%
%
%
 
%
%
\end{document}